\documentclass{fac}
\usepackage{graphicx}
\usepackage{amssymb}
\usepackage{amsmath}
\ifprodtf \else \usepackage{latexsym}\fi

\newcommand\black{\ensuremath{\blacktriangleright}}
\newcommand\white{\ensuremath{\vartriangleright}}

\newif\ifamsfontsloaded
\ifprodtf
  \newcommand\whbl{\white\kern-.1em--\kern-.1em\black}
  \newcommand\blwh{\black\kern-.1em--\kern-.1em\white}
  \newcommand\blbl{\black\kern-.1em--\kern-.1em\black}
  \newcommand\whwh{\white\kern-.1em--\kern-.1em\white}
  \amsfontsloadedtrue
\else
  \checkfont{msam10}
  \iffontfound
    \IfFileExists{amssymb.sty}
      {\usepackage{amssymb}\amsfontsloadedtrue
       \newcommand\whbl{\white\kern-.125em--\kern-.125em\black}%
       \newcommand\blwh{\black\kern-.125em--\kern-.125em\white}%
       \newcommand\blbl{\black\kern-.125em--\kern-.125em\black}%
       \newcommand\whwh{\white\kern-.125em--\kern-.125em\white}}
      {}
  \fi
\fi

\title[A Survey on Formal Methods for Web Service Composition]
      {A Survey on Formal Methods for Web Service Composition}

\author[Yong Wang]
    {Yong Wang\\
     College of Computer Science and Technology,\\
     Beijing University of Technology, Beijing, China\\
     }

\correspond{Yong Wang, Pingleyuan 100, Chaoyang District, Beijing, China.
            e-mail: wangy@bjut.edu.cn}

\pubyear{2011}
\pagerange{\pageref{firstpage}--\pageref{lastpage}}

\begin{document}
\label{firstpage}

\makecorrespond

\maketitle

\begin{abstract}
Web Service Composition creates new composite Web Services from existing Web Services which embodies the added values of Web Service technology and is a key technology to solve cross-organizational business process integrations. We do a survey on formal methods for Web Service Composition in the following way. Through analyses of Web Service Composition, we establish a reference model called RM-WSComposition to capture elements of Web Service Composition. Based on the RM-WSComposition, issues on formalization for Web Service Composition are pointed out and state-of-the-art on formal methods for Web Service Composition is introduced. Finally, we point out the trends on this topic. For convenience, we use an example called BuyingBooks to illustrate the concepts and mechanisms in Web Service Composition.
\end{abstract}


\section{Introduction}

Web Service (WS) is a quite new distributed software component which uses the Web as its provision platform. That is, like other distributed components, such as DCOM, EJB, CORBA, etc, a WS has its interface described by WSDL\cite{WSDL}, its communication protocol named SOAP\cite{SOAP} based on HTTP, and its name and directory service called UDDI\cite{UDDI}.

WS aims at the requirements of cross-organizational business integration, such as e-commerce, and is developed with a multi-layer protocol stack which models different aspects of cross-organizational business integration based on WSes, including long running business transactions, security, QoS, coordination, etc, and also the so-called WS Composition.

In a viewpoint of requirements of cross-organizational business integration, WS Composition is trying to solve the problems of business process integration, including modeling of business processes and their interactions, and uses WSes as basic function units to model business activities. In a view of WS technologies, WS Composition creates new composite WSes from the set of existing WSes through some composition patterns.

WS Composition usually utilizes two kind of composition patterns\cite{WSOandWSC1}, one is a workflow-like\cite{WFMCRM} pattern called Web Service Orchestration (WSO) to model business processes, and the other is an aggregate pattern named Web Service Choreoraphy (WSC) to model interactions among business processes\footnote[1]{Strictly, a WSC is not dependent on WSOs. It can also model interactions among WSes which are not implemented as WSOs}.

In research on WS Composition, there are three way of efforts. The first is the industry way which tries to establish uniform specifications of WS Composition, such as the WSO specification WS-BPEL\cite{WS-BPEL} and the WSC specification WS-CDL\cite{WS-CDL}. The second pursues automatic composition of WSes\cite{SemanticWSComposition}\cite{SemanticWSComposition2}, such as efforts of OWL\cite{OWL} based on semantic Web technology. And the last one is formal methods of WS Composition which tries to give WS Compostion a firm theoretical foundation based on different formal tools.

In this paper, we do a survey on formal methods for WS Composition. Based on analyses of requirements for WS Composition, we establish the reference model called RM-WSComposition to capture the elements of WS Composition and point out issues on formalization for WS Composition. Then we introduce state-of-the-art of formalization for WS Composition. Finally, we conclude the research works and point out the trend. Throughout the paper, we use a example called BuyingBooks to illustrate the concepts and mechanisms in the related works.

\section{An Example of Cross-Organizational Business Process Integration, BuyingBooks}\label{BuyingBooksExample}

\begin{figure}
  \centering
  %\vspace{5cm}
  \includegraphics{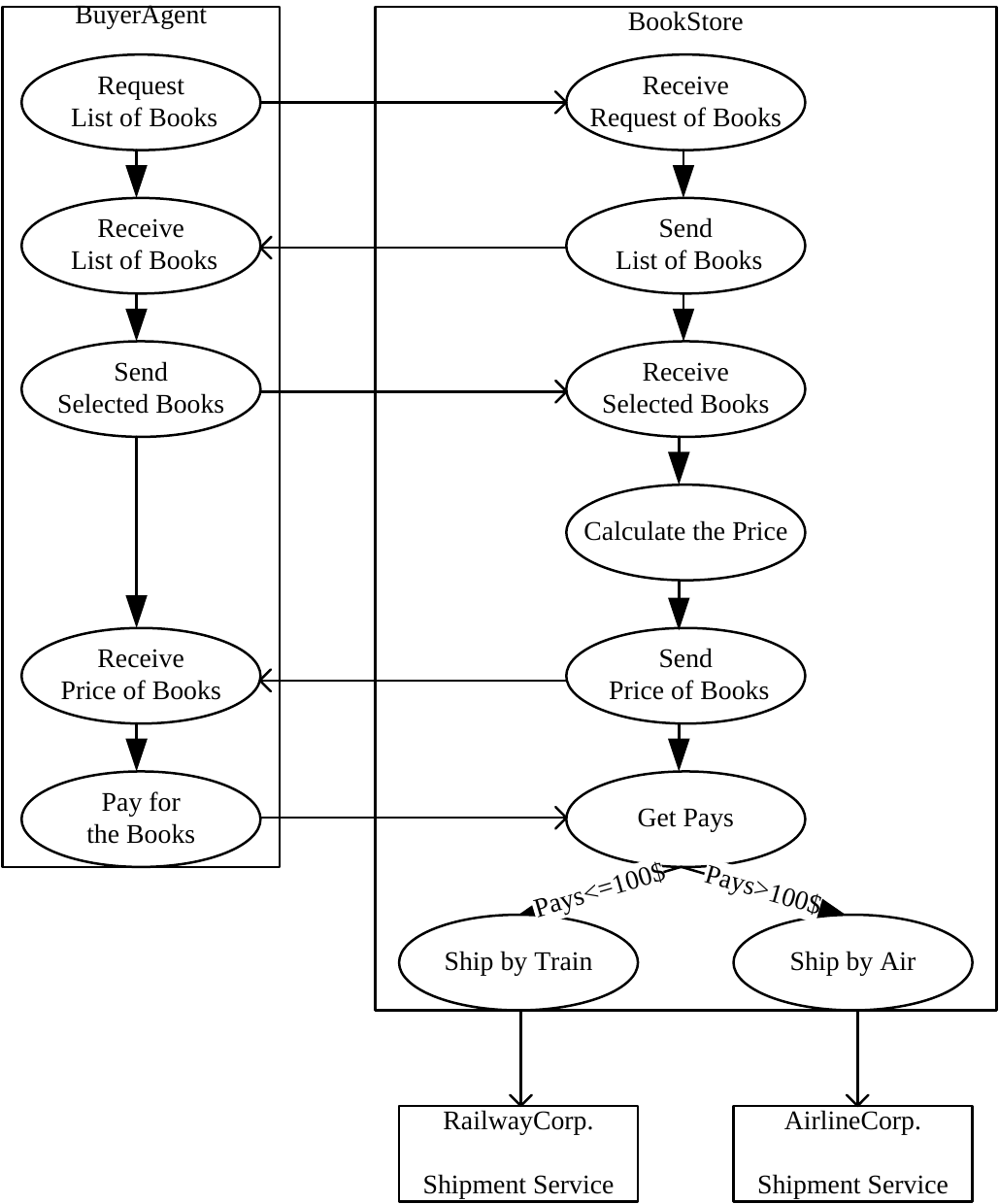}
  \caption{The BuyingBooks Example.}
  \label{Fig.BuyingBooksExample}
\end{figure}

Now, we give an example of cross-organizational business process integration called BuyingBooks as Fig.\ref{Fig.BuyingBooksExample} shows. We use this BuyingBooks example throughout this paper to illustrate concepts and mechanisms in WS Composition and its formal methods.

In Fig.\ref{Fig.BuyingBooksExample}, there are four organizations: BuyerAgent, BookStore, RailwayCorp, and AirlineCorp. And each organization has one business process. Exactly, there are two business processes, the business processes in RailwayCorp and AirlineCorp are simplified as just WSes for simpleness without loss of generality. We introduce the business process of BookStore as follows, and the process of BuyerAgent can be gotten in contrary.

\begin{enumerate}
  \item The BookStore receives request of list of books from the buyer through BuyerAgent.
  \item It sends the list of books to the buyer via BuyerAgent.
  \item It receives the selected book list by the buyer via BuyerAgent.
  \item Then it calculates the price of the selected books.
  \item It sends the price of the selected books to the buyer via BuyerAgent.
  \item Then it gets pays for the books from the buyer via BuyerAgent.
  \item If the pays are greater than 100\$, then the BookStore calls the shipment service of AirlineCorp for the shipment of books.
  \item Else, the BookStore calls the shipment service of RailwayCorp for the shipment of book. Then the process is ended.
\end{enumerate}

\section{A Reference Model of Web Service Composition}

In this section, we deeply discuss some topics on WS Composition. Based on these discussions, we introduce a reference model of WS Composition called RM-WSComposition.

\subsection{Web Service Composition and Cross-Organizational Workflow}

Workflow\cite{WFMCRM}\cite{WFMCPD} is also a solution to cross-organizational business process integration. Workflow in such situation is called cross-organizational workflow\cite{InterorgaWorkflowPNet}\cite{InterorgaWorkfowPView1}\cite{InterorgaWorkfowPView2}. But the implementations of activities in a cross-organizational workflow are not limited to WSes.

Aalst uses Petri Net to model a workflow called WF-Net\cite{WF-Net} and connects two WF-Nets within different organizations into a newly global WF-Net to model integration of two workflow. An obvious problem is that a partner workflow is located in the inner of an organization, that is, the details of an inner workflow are hidden to the external world, so a global view of entire connection with two detailed inner workflow usually can not be gotten.

This leads to emergence of the so-called process view\cite{InterorgaWorkfowPView2} and integration of two inner process can be implemented based on process views\cite{InterorgaWorkfowPView1}. A process view is an observable version of an inner process from outside and serves as the interface of an inner process.

Research efforts on cross-organizational workflow give a great reference to WS Composition, especially WSO. WSOs aimed at cross-organizational business process integrations and used a workflow-like pattern to orchestrate WSes are surely some kinds cross-organizational workflows within which have activities implemented by WSes outside.

But, are mechanisms on cross-organizational workflow, such as process view, enough for those of WS Composition? We will answer this question in the background of WS technology in the following.

\subsection{Web Service Orchestration and Its Interface}

As discussed above, a WSO usually is a cross-organizational workflow to orchestrate WSes and also is encapsulated as a WS to receive messages from outside. In this section, we will discuss about WSO and its interface.

\subsubsection{Web Service Orchestration}

The main efforts on WSO of the industry are trying to establish a uniform WSO description language specification, such as the early WSFL\cite{WSFL}, XLANG\cite{XLANG}, and lately converged WS-BPEL\cite{WS-BPEL}. Such WSO description languages based on different mathematic models have constructs to model invocation of WSes, manipulate information transferred between WSes, control execution flows of these activities and inner transaction processing mechanisms. WSO description language can be used to define various WSOs under different requirements and acts as a so-called meta language.

A WSO description language, such as WS-BPEL, has:

 \begin{itemize}
   \item basic constructs called atomic activities to model invocation to a WS outside, receiving messages from a WS outside and reply to that WS, and other inner basic functions;
   \item information and variables exchanged between WSes;
   \item control flows called structural activities to orchestrate activities;
   \item other inner transaction processing mechanisms, such as exception definitions and throwing mechanisms, event definitions and response mechanisms.
 \end{itemize}

A WSO described by such a meta language is enabled by the meta language interpreter called WSO engine. A WSO is hidden in the inner of an organization and only can interact with outside through its interface. Since a WSO is also encapsulated as a WS, its WS can act as the interface of the WSO.

The BookStore WSO described by WS-BPEL is given as follows.

-------------------------------------------------------------------------------

$\langle$process name="BookStore"

\quad targetNamespace="http://example.wscs.com/2011/ws-bp/bookstore"

\quad xmlns="http://docs.oasis-open.org/wsbpel/2.0/process/executable"

\quad xmlns:bns="http://example.wscs.com/2011/wsdl/BuyerAgent.wsdl"

\quad xmlns:ans="http://example.wscs.com/2011/wsdl/AirlineCorp.wsdl"

\quad xmlns:rns="http://example.wscs.com/2011/wsdl/RailwayCorp.wsdl"

\quad xmlns:lns="http://example.wscs.com/2011/wsdl/BookStore.wsdl"$\rangle$

\quad $\langle$documentation xml:lang="EN"$\rangle$

\quad\quad This document describes the BookStore process.

\quad $\langle$/documentation$\rangle$

\quad $\langle$partnerLinks$\rangle$

\quad\quad $\langle$partnerLink name="bookStoreAndBuyerAgent"

\quad\quad\quad partnerLinkType="lns:bookStoreAndBuyerAgentLT"

\quad\quad\quad myRole="bookStore" partnerRole="buyerAgent" /$\rangle$

\quad\quad $\langle$partnerLink name="bookStoreAndRailwayCorp"

\quad\quad\quad partnerLinkType="lns:bookStoreAndRailwayCorpLT"

\quad\quad\quad myRole="bookStore" partnerRole="railwayCorp" /$\rangle$

\quad\quad $\langle$partnerLink name="bookStoreAndAirlineCorp"

\quad\quad\quad partnerLinkType="lns:bookStoreAndAirlineCorpLT"

\quad\quad\quad myRole="bookStore" partnerRole="airlineCorp" /$\rangle$

\quad $\langle$/partnerLinks$\rangle$

\quad $\langle$variables$\rangle$

\quad\quad $\langle$variable name="RequestListofBooks" messageType="lns:requestListofBooks"/$\rangle$

\quad\quad $\langle$variable name="RequestListofBooksResponse" messageType="lns:requestListofBooksResponse"/$\rangle$

\quad\quad $\langle$variable name="ListofBooks" messageType="lns:listofBooks"/$\rangle$

\quad\quad $\langle$variable name="ListofBooksResponse" messageType="lns:listofBooksResponse"/$\rangle$

\quad\quad $\langle$variable name="SelectListofBooks"  messageType="lns:selectListofBooks"/$\rangle$

\quad\quad $\langle$variable name="SelectListofBooksResponse"  messageType="lns:selectListofBooksResponse"/$\rangle$

\quad\quad $\langle$variable name="Price" messageType="lns:price"/$\rangle$

\quad\quad $\langle$variable name="PriceResponse" messageType="lns:priceResponse"/$\rangle$

\quad\quad $\langle$variable name="Pays" messageType="lns:pays"/$\rangle$

\quad\quad $\langle$variable name="PaysResponse" messageType="lns:paysResponse"/$\rangle$

\quad\quad $\langle$variable name="ShipmentByTrain" messageType="lns:shipmentByTrain"/$\rangle$

\quad\quad $\langle$variable name="ShipmentByTrainResponse" messageType="lns:shipmentByTrainResponse"/$\rangle$

\quad\quad $\langle$variable name="ShipmentByAir" messageType="lns:shipmentByAir"/$\rangle$

\quad\quad $\langle$variable name="ShipmentByAirResponse" messageType="lns:shipmentByAirResponse"/$\rangle$

\quad $\langle$/variables$\rangle$

\quad $\langle$sequence$\rangle$

\quad\quad $\langle$receive partnerLink="bookStoreAndBuyerAgent"

\quad\quad\quad portType="lns:bookStore4BuyerAgentInterface"

\quad\quad\quad operation="opRequestListofBooks" variable="RequestListofBooks"

\quad\quad\quad createInstance="yes"$\rangle$

\quad\quad $\langle$/receive$\rangle$

\quad\quad $\langle$invoke partnerLink="bookStoreAndBuyerAgent"

\quad\quad\quad portType="bns:buyAgent4BookStoreInterface"

\quad\quad\quad operation="opReceiveListofBooks" inputVariable="ListofBooks"

\quad\quad\quad outputVariable="ListofBooksResponse"$\rangle$

\quad\quad $\langle$/invoke$\rangle$

\quad\quad $\langle$receive partnerLink="bookStoreAndBuyerAgent"

\quad\quad\quad portType="lns:bookStore4BuyerAgentInterface"

\quad\quad\quad operation="opSelectListofBooks" variable="SelectListofBooks"$\rangle$

\quad\quad $\langle$/receive$\rangle$

\quad\quad $\langle$reply partnerLink="bookStoreAndBuyerAgent"

\quad\quad\quad portType="lns:bookStore4BuyerAgentInterface"

\quad\quad\quad operation="opSelectListofBooks" variable="SelectListofBooksResponse"$\rangle$

\quad\quad $\langle$/reply$\rangle$

\quad\quad $\langle$!--inner activity: calculate the price of selected books--$\rangle$

\quad\quad $\langle$invoke partnerLink="bookStoreAndBuyerAgent"

\quad\quad\quad portType="bns:buyAgent4BookStoreInterface"

\quad\quad\quad operation="opReceivePrice" inputVariable="Price"

\quad\quad\quad outputVariable="PriceResponse"$\rangle$

\quad\quad $\langle$receive partnerLink="bookStoreAndBuyerAgent"

\quad\quad\quad portType="lns:bookStore4BuyerAgentInterface"

\quad\quad\quad operation="opPays" variable="Pays"$\rangle$

\quad\quad $\langle$/receive$\rangle$

\quad\quad $\langle$reply partnerLink="bookStoreAndBuyerAgent"

\quad\quad\quad portType="lns:bookStore4BuyerAgentInterface"

\quad\quad\quad operation="opPays" variable="PaysResponse"$\rangle$

\quad\quad $\langle$if$\rangle$

\quad\quad\quad$\langle$condition$\rangle$ getVariable('Price') $\langle$ 100 $\langle$/condition$\rangle$

\quad\quad\quad $\langle$invoke partnerLink="bookStoreAndAirlineCorp"

\quad\quad\quad\quad portType="ans:airlineCorp4BookStoreInterface"

\quad\quad\quad\quad operation="opShipmentByAir" inputVariable="ShipmentByAir"

\quad\quad\quad\quad outputVariable="ShipmentByAirResponse"$\rangle$

\quad\quad\quad $\langle$else$\rangle$

\quad\quad\quad\quad $\langle$invoke partnerLink="bookStoreAndRailwayCorp"

\quad\quad\quad\quad\quad portType="rns:railwayCorp4BookStoreInterface"

\quad\quad\quad\quad\quad operation="opShipmentByTrain" inputVariable="ShipmentByTrain"

\quad\quad\quad\quad\quad outputVariable="ShipmentByTrainResponse"$\rangle$

\quad\quad\quad $\langle$/else$\rangle$

\quad\quad $\langle$/if$\rangle$

\quad $\langle$/sequence$\rangle$

$\langle$/process$\rangle$

-------------------------------------------------------------------------------

\subsubsection{Stateless Web Service and Stateful Web Service}

In the viewpoint of W3C, a WS itself is an interface or a wrapper of an application inside the boundary of an organization that has a willing to interact with applications outside. That is, a W3C WS has no an independent programming model like other component models and has no needs of containing local states for local computations. Indeed, there are different sounds of developing WS to be a full sense component, Such as OGSI\cite{OGSI}. Incompatibility between W3C WS and OGSI-like WS leads to WSRF\cite{WSRF} as a compromised solution which reserves the W3C WS and develops a notion of WS Resource to model states.

Since interactions among WSes are eventually driven by their inner applications, at runtime, there is no any necessary to maintain states for a WS. That is, a WS just \emph{deliver}\footnote[1]{without any processing} messages, including incoming messages and outgoing messages.

The BookStore WS described by WSDL is following.

-------------------------------------------------------------------------------

$\langle$?xml version="1.0" encoding="utf-8"?$\rangle$

$\langle$description

\quad xmlns="http://www.w3.org/2004/08/wsdl"

\quad\quad\quad xmlns:plnk="http://docs.oasis-open.org/wsbpel/2.0/plnktype"

\quad targetNamespace= "http://example.wscs.com/2011/wsdl/BookStore.wsdl"

\quad xmlns:tns= "http://example.wscs.com/2011/wsdl/BookStore.wsdl"

\quad xmlns:ghns = "http://example.wscs.com/2011/schemas/BookStore.xsd"

\quad xmlns:bans = "http://example.wscs.com/2011/wsdl/BuyerAgent.wsdl"

\quad xmlns:rcns = "http://example.wscs.com/2011/wsdl/RailwayCorp.wsdl"

\quad xmlns:acns = "http://example.wscs.com/2011/wsdl/AirlineCorp.wsdl"

\quad xmlns:wsoap= "http://www.w3.org/2004/08/wsdl/soap12"

\quad xmlns:soap="http://www.w3.org/2003/05/soap-envelope"$\rangle$

\quad $\langle$documentation$\rangle$

\quad \quad This document describes the BookStore Web service.

\quad $\langle$/documentation$\rangle$

\quad$\langle$types$\rangle$

\quad\quad$\langle$xs:schema

\quad\quad\quad xmlns:xs="http://www.w3.org/2001/XMLSchema"

\quad\quad\quad targetNamespace="http://example.wscs.com/2011/schemas/BookStore.xsd"

\quad\quad\quad xmlns="http://example.wscs.com/2011/schemas/BookStore.xsd"$\rangle$

\quad\quad\quad $\langle$xs:element name="requestListofBooks" type="tRequestListofBooks"/$\rangle$

\quad\quad\quad $\langle$xs:complexType name="tRequestListofBooks"/$\rangle$

\quad\quad\quad $\langle$xs:element name="requestListofBooksReponse"

\quad\quad\quad\quad type="tRequestListofBooksResponse"/$\rangle$

\quad\quad\quad $\langle$xs:complexType name="tRequestListofBooksResponse"/$\rangle$

\quad\quad\quad $\langle$xs:element name="listofBooks" type="tListofBooks"/$\rangle$

\quad\quad\quad $\langle$xs:complexType name="tListofBooks"/$\rangle$

\quad\quad\quad $\langle$xs:element name="listofBooksResponse" type="tListofBooksResponse"/$\rangle$

\quad\quad\quad $\langle$xs:complexType name="tListofBooksResponse"/$\rangle$

\quad\quad\quad $\langle$xs:element name="selectListofBooks" type="tSelectListofBooks"/$\rangle$

\quad\quad\quad $\langle$xs:complexType name="tSelectListofBooks"/$\rangle$

\quad\quad\quad $\langle$xs:element name="selectListofBooksResponse"

\quad\quad\quad\quad type="tSelectListofBooksResponse"/$\rangle$

\quad\quad\quad $\langle$xs:complexType name="tSelectListofBooksResponse"/$\rangle$

\quad\quad\quad $\langle$xs:element name="price" type="xs:float"/$\rangle$

\quad\quad\quad $\langle$xs:element name="priceResponse" type="tPriceResponse"/$\rangle$

\quad\quad\quad $\langle$xs:complexType name="tPriceResponse"/$\rangle$

\quad\quad\quad $\langle$xs:element name="pays" type="tPays"/$\rangle$

\quad\quad\quad $\langle$xs:complexType name="tPays"/$\rangle$

\quad\quad\quad $\langle$xs:element name="paysResponse" type="tPaysResponse"/$\rangle$

\quad\quad\quad $\langle$xs:complexType name="tPaysResponse"/$\rangle$

\quad\quad\quad $\langle$xs:element name="shipmentByTrain" type="tShipmentByTrain"/$\rangle$

\quad\quad\quad $\langle$xs:complexType name="tShipmentByTrain"/$\rangle$

\quad\quad\quad $\langle$xs:element name="shipmentByTrainResponse"

\quad\quad\quad\quad type="tShipmentByTrainResponse"/$\rangle$

\quad\quad\quad $\langle$xs:complexType name="tShipmentByTrainResponse"/$\rangle$

\quad\quad\quad $\langle$xs:element name="shipmentByAir" type="tShipmentByAir"/$\rangle$

\quad\quad\quad $\langle$xs:complexType name="tShipmentByAir"/$\rangle$

\quad\quad\quad $\langle$xs:element name="shipmentByAirResponse" type="tShipmentByAirResponse"/$\rangle$

\quad\quad\quad $\langle$xs:complexType name="tShipmentByAirResponse"/$\rangle$

\quad\quad $\langle$/xs:schema$\rangle$

\quad $\langle$/types$\rangle$

\quad $\langle$interface name = "bookStore4BuyerAgentInterface"$\rangle$

\quad\quad$\langle$operation name="opRequestListofBooks"$\rangle$

\quad\quad\quad $\langle$input messageLabel="InOpRequestListofBooks"

\quad\quad\quad\quad element="ghns:requestListofBooks" /$\rangle$

\quad\quad\quad $\langle$output messageLabel="OutOpRequestListofBooks"

\quad\quad\quad\quad element="ghns:requestListofBooksReponse" /$\rangle$

\quad\quad $\langle$/operation$\rangle$

\quad\quad $\langle$operation name="opSelectListofBooks"$\rangle$

\quad\quad\quad $\langle$input messageLabel="InOpSelectListofBooks"

\quad\quad\quad\quad element="ghns:selectListofBooks" /$\rangle$

\quad\quad\quad $\langle$output messageLabel="OutOpSelectListofBooks"

\quad\quad\quad\quad element="ghns:selectListofBooksResponse" /$\rangle$

\quad\quad $\langle$/operation$\rangle$

\quad\quad $\langle$operation name="opPays"$\rangle$

\quad\quad\quad $\langle$input messageLabel="InOpPays"

\quad\quad\quad\quad element="ghns:pays" /$\rangle$

\quad\quad\quad $\langle$output messageLabel="OutOpPays"

\quad\quad\quad\quad element="ghns:paysResponse" /$\rangle$

\quad\quad $\langle$/operation$\rangle$

\quad $\langle$/interface$\rangle$

\quad $\langle$plnk:partnerLinkType name="bookStoreAndBuyerAgentLT"$\rangle$

\quad\quad $\langle$plnk:role name="bookStore"

\quad\quad\quad portType="tns:bookStore4BuyerAgentInterface" /$\rangle$

\quad\quad $\langle$plnk:role name="buyerAgent"

\quad\quad\quad portType="bans:buyerAgent4BookStoreInterface" /$\rangle$

\quad $\langle$/plnk:partnerLinkType$\rangle$

\quad $\langle$plnk:partnerLinkType name="bookStoreAndRaiwayCorpLT"$\rangle$

\quad\quad $\langle$plnk:role name="railwayCorp"

\quad\quad\quad portType="rcns:railwayCorp4BookStoreInterface" /$\rangle$

\quad $\langle$/plnk:partnerLinkType$\rangle$

\quad $\langle$plnk:partnerLinkType name="bookStoreAndAirlineCorpLT"$\rangle$

\quad\quad $\langle$plnk:role name="airlineCorp"

\quad\quad\quad portType="acns:airlineCorp4BookStoreInterface" /$\rangle$

\quad $\langle$/plnk:partnerLinkType$\rangle$

$\langle$/description$\rangle$

-------------------------------------------------------------------------------

\subsubsection{Is WSRF-Style Web Service Enough?}

As a runtime mechanism, WSRF-style WS can serve as an interface of a WSO enough. But, a WSO, even a WS Resource, is surely different to a stateless function for share. That is, from a view of an observer outside, the interfaces\footnote[1]{exactly WSDL Operations of an interface WS of a WSO} of a WSO surely are stateful and must be invoked under the constraints of the WSO. Note that, different observers may have different views to the same WSO.

\begin{figure}
  \centering
  %\vspace{5cm}
  \includegraphics{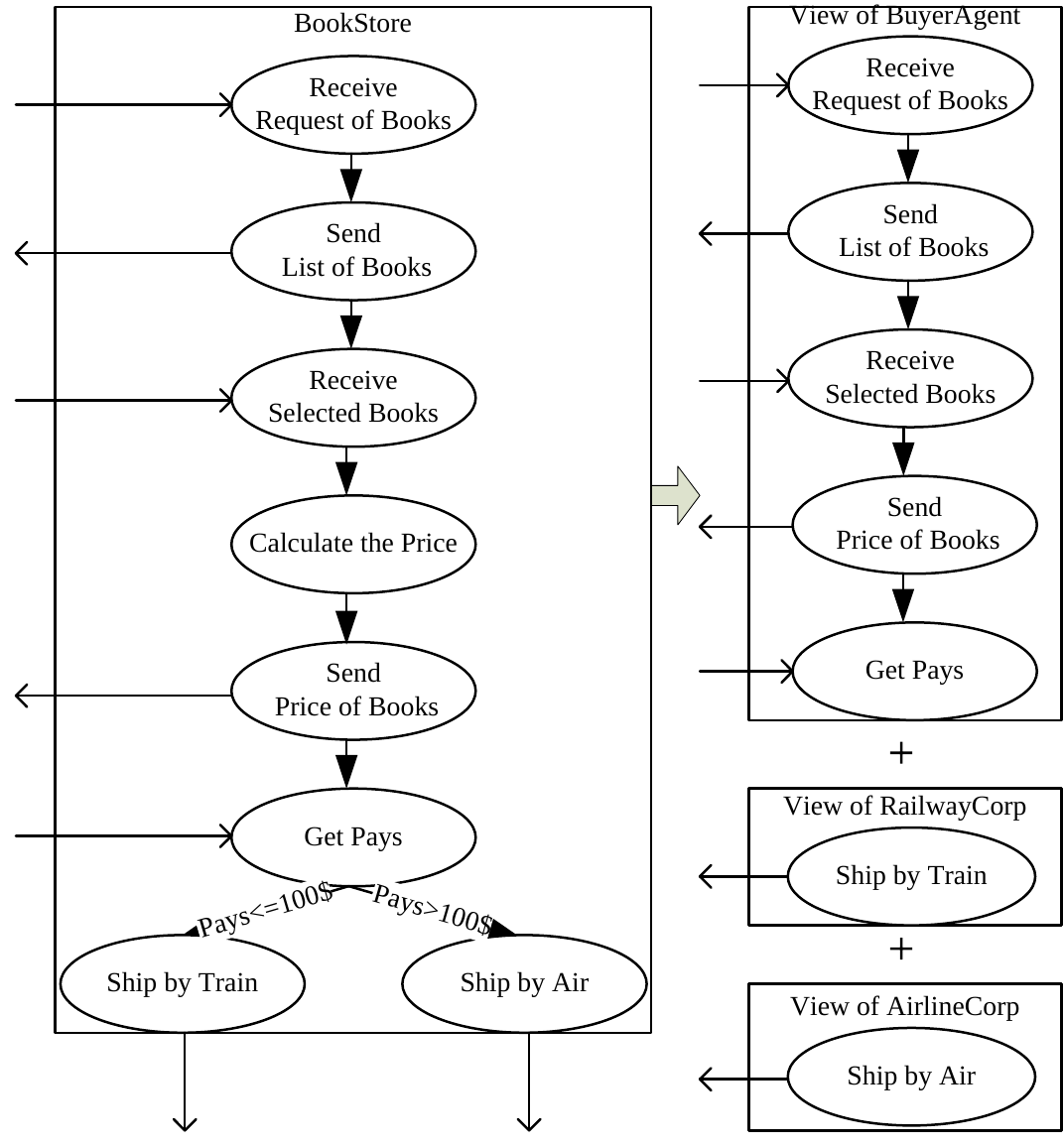}
  \caption{Views of BookStore Process from Different Observers.}
  \label{Fig.Views}
\end{figure}

BuyerAgent, RailwayCorp and AirlineCorp have different views to the same BookStore WSO as Fig.\ref{Fig.Views} shows. Note that there is no an inner "Calculate the Price" activity in the observation of BuyerAgent.

Such observable views of a WSO are called abstract processes in WS-BPEL\cite{WS-BPEL} and are called WSC Interfaces in WSCI\cite{WSCI}. But these specifications do not explain their functions and even the relations to the original WSO. Though a WSRF-style WS can act as the interfaces of a WSO at runtime, but it has few uses at design time. Since observable views have more information about the inner WSO, they can be used to do more things at design time, such as generation of a WSC and generation of stub codes being more expressive than those of WSDL.

That is, observable views of a WSO are more expressive interface descriptions than WSDL. The information of observable views can be available by extending WSDL to be annotations of WSDL interfaces.

The BookStore4BuyerAgent observable view described by abstract process of WS-BPEL is as follows.

-------------------------------------------------------------------------------

$\langle$process name="BookStoreAbstract4BuyerAgent"

\quad targetNamespace="http://example.wscs.com/2011/ws-bp/bookstore"

\quad xmlns="http://docs.oasis-open.org/wsbpel/2.0/process/abstract"

\quad abstractProcessProfile="http://docs.oasis-open.org/wsbpel/2.0/process/abstract/ap11/2006/08"

\quad xmlns:bns="http://example.wscs.com/2011/wsdl/BuyerAgent.wsdl"

\quad xmlns:lns="http://example.wscs.com/2011/wsdl/BookStore.wsdl"$\rangle$

\quad $\langle$documentation xml:lang="EN"$\rangle$

\quad\quad This document describes the BookStore abstract process for BuyerAgent.

\quad $\langle$/documentation$\rangle$

\quad $\langle$partnerLinks$\rangle$

\quad\quad $\langle$partnerLink name="bookStoreAndBuyerAgent"

\quad\quad\quad partnerLinkType="lns:bookStoreAndBuyerAgentLT"

\quad\quad\quad myRole="bookStore" partnerRole="buyerAgent" /$\rangle$

\quad $\langle$/partnerLinks$\rangle$

\quad $\langle$variables$\rangle$

\quad\quad $\langle$variable name="RequestListofBooks" messageType="lns:requestListofBooks"/$\rangle$

\quad\quad $\langle$variable name="RequestListofBooksResponse" messageType="lns:requestListofBooksResponse"/$\rangle$

\quad\quad $\langle$variable name="ListofBooks" messageType="lns:listofBooks"/$\rangle$

\quad\quad $\langle$variable name="ListofBooksResponse" messageType="lns:listofBooksResponse"/$\rangle$

\quad\quad $\langle$variable name="SelectListofBooks"  messageType="lns:selectListofBooks"/$\rangle$

\quad\quad $\langle$variable name="SelectListofBooksResponse"  messageType="lns:selectListofBooksResponse"/$\rangle$

\quad\quad $\langle$variable name="Price" messageType="lns:price"/$\rangle$

\quad\quad $\langle$variable name="PriceResponse" messageType="lns:priceResponse"/$\rangle$

\quad\quad $\langle$variable name="Pays" messageType="lns:pays"/$\rangle$

\quad\quad $\langle$variable name="PaysResponse" messageType="lns:paysResponse"/$\rangle$

\quad $\langle$/variables$\rangle$

\quad $\langle$sequence$\rangle$

\quad\quad $\langle$receive partnerLink="bookStoreAndBuyerAgent"

\quad\quad\quad portType="lns:bookStore4BuyerAgentInterface"

\quad\quad\quad operation="opRequestListofBooks" variable="RequestListofBooks"

\quad\quad\quad createInstance="yes"$\rangle$

\quad\quad $\langle$/receive$\rangle$

\quad\quad $\langle$invoke partnerLink="bookStoreAndBuyerAgent"

\quad\quad\quad portType="bns:buyAgent4BookStoreInterface"

\quad\quad\quad operation="opReceiveListofBooks" inputVariable="ListofBooks"

\quad\quad\quad outputVariable="ListofBooksResponse"$\rangle$

\quad\quad $\langle$/invoke$\rangle$

\quad\quad $\langle$receive partnerLink="bookStoreAndBuyerAgent"

\quad\quad\quad portType="lns:bookStore4BuyerAgentInterface"

\quad\quad\quad operation="opSelectListofBooks" variable="SelectListofBooks"$\rangle$

\quad\quad $\langle$/receive$\rangle$

\quad\quad $\langle$reply partnerLink="bookStoreAndBuyerAgent"

\quad\quad\quad portType="lns:bookStore4BuyerAgentInterface"

\quad\quad\quad operation="opSelectListofBooks" variable="SelectListofBooksResponse"$\rangle$

\quad\quad $\langle$/reply$\rangle$

\quad\quad $\langle$!--inner activity: calculate the price of selected books--$\rangle$

\quad\quad $\langle$invoke partnerLink="bookStoreAndBuyerAgent"

\quad\quad\quad portType="bns:buyAgent4BookStoreInterface"

\quad\quad\quad operation="opReceivePrice" inputVariable="Price"

\quad\quad\quad outputVariable="PriceResponse"$\rangle$

\quad\quad $\langle$receive partnerLink="bookStoreAndBuyerAgent"

\quad\quad\quad portType="lns:bookStore4BuyerAgentInterface"

\quad\quad\quad operation="opPays" variable="Pays"$\rangle$

\quad\quad $\langle$/receive$\rangle$

\quad\quad $\langle$reply partnerLink="bookStoreAndBuyerAgent"

\quad\quad\quad portType="lns:bookStore4BuyerAgentInterface"

\quad\quad\quad operation="opPays" variable="PaysResponse"$\rangle$

\quad $\langle$/sequence$\rangle$

$\langle$/process$\rangle$

-------------------------------------------------------------------------------

And the BookStore4RailwayCorp observable view described by abstract process of WS-BPEL is as follows.

-------------------------------------------------------------------------------

$\langle$process name="BookStoreAbstract4RailwayCorp"

\quad targetNamespace="http://example.wscs.com/2011/ws-bp/bookstore"

\quad abstractProcessProfile="http://docs.oasis-open.org/wsbpel/2.0/process/abstract/ap11/2006/08"

\quad xmlns="http://docs.oasis-open.org/wsbpel/2.0/process/abstract"

\quad xmlns:rns="http://example.wscs.com/2011/wsdl/RailwayCorp.wsdl"

\quad xmlns:lns="http://example.wscs.com/2011/wsdl/BookStore.wsdl"$\rangle$

\quad $\langle$documentation xml:lang="EN"$\rangle$

\quad\quad This document describes the BookStore abstract process for RailwayCorp.

\quad $\langle$/documentation$\rangle$

\quad $\langle$partnerLinks$\rangle$

\quad\quad $\langle$partnerLink name="bookStoreAndRailwayCorp"

\quad\quad\quad partnerLinkType="lns:bookStoreAndRailwayCorpLT"

\quad\quad\quad myRole="bookStore" partnerRole="railwayCorp" /$\rangle$

\quad $\langle$/partnerLinks$\rangle$

\quad $\langle$variables$\rangle$

\quad\quad $\langle$variable name="ShipmentByTrain" messageType="lns:shipmentByTrain"/$\rangle$

\quad\quad $\langle$variable name="ShipmentByTrainResponse" messageType="lns:shipmentByTrainResponse"/$\rangle$

\quad $\langle$/variables$\rangle$

\quad $\langle$invoke partnerLink="bookStoreAndRailwayCorp"

\quad\quad portType="rns:railwayCorp4BookStoreInterface"

\quad\quad operation="opShipmentByTrain" inputVariable="ShipmentByTrain"

\quad\quad outputVariable="ShipmentByTrainResponse"$\rangle$

$\langle$/process$\rangle$

-------------------------------------------------------------------------------

At last, the BookStore4AirlineCorp observable view described by abstract process of WS-BPEL is following.

-------------------------------------------------------------------------------

$\langle$process name="BookStoreAbstract4AirlineCorp"

\quad targetNamespace="http://example.wscs.com/2011/ws-bp/bookstore"

\quad abstractProcessProfile="http://docs.oasis-open.org/wsbpel/2.0/process/abstract/ap11/2006/08"

\quad xmlns="http://docs.oasis-open.org/wsbpel/2.0/process/abstract"

\quad xmlns:rns="http://example.wscs.com/2011/wsdl/AirlineCorp.wsdl"

\quad xmlns:lns="http://example.wscs.com/2011/wsdl/BookStore.wsdl"$\rangle$

\quad $\langle$documentation xml:lang="EN"$\rangle$

\quad\quad This document describes the BookStore abstract process for AirlineCorp.

\quad $\langle$/documentation$\rangle$

\quad $\langle$partnerLinks$\rangle$

\quad\quad $\langle$partnerLink name="bookStoreAndAirlineCorp"

\quad\quad\quad partnerLinkType="lns:bookStoreAndAirlineCorpLT"

\quad\quad\quad myRole="bookStore" partnerRole="airlineCorp" /$\rangle$

\quad $\langle$/partnerLinks$\rangle$

\quad $\langle$variables$\rangle$

\quad\quad $\langle$variable name="ShipmentByAir" messageType="lns:shipmentByAir"/$\rangle$

\quad\quad $\langle$variable name="ShipmentByAirResponse" messageType="lns:shipmentByAirResponse"/$\rangle$

\quad $\langle$/variables$\rangle$

\quad $\langle$invoke partnerLink="bookStoreAndAirlineCorp"

\quad\quad portType="ans:airlineCorp4BookStoreInterface"

\quad\quad operation="opShipmentByAir" inputVariable="ShipmentByAir"

\quad\quad outputVariable="ShipmentByAirResponse"$\rangle$

$\langle$/process$\rangle$

-------------------------------------------------------------------------------

\subsection{Web Service Choreography and Its Enablement}

A WSC acts as a contract or a protocol among interacting WSes. In a viewpoint of business integration requirements, a WSC serves as a business contract to constrain the rights and obligations of business partners. And from a view of WS technology, we can treat a WSC as a communication protocol which coordinates the interaction behaviors of involved WSes.

The main elements of a WSC defined by a WSC description language, such as WS-CDL\cite{WS-CDL}, are following:

\begin{itemize}
  \item partners within a WSC including the role of a partner acting as and relationships among partners;
  \item local states exchanged among the interacting WSes;
  \item interaction points and interaction behaviors defined as the core contents in a WSC;
  \item sequence definitions of interaction behaviors;
  \item other inner transaction processing mechanisms, such as exception definitions and throwing mechanisms, event definitions and response mechanisms.
\end{itemize}

The BuyerAgentAndBookStore WSC described by WS-CDL is as follows.

-------------------------------------------------------------------------------

$\langle$?xml version="1.0" encoding="UTF-8"?$\rangle$

$\langle$package xmlns="http://www.w3.org/2005/10/cdl"

\quad xmlns:cdl="http://www.w3.org/2005/10/cdl"

\quad xmlns:xsi="http://www.w3.org/2001/XMLSchema-instance"

\quad xmlns:xsd="http://www.w3.org/2001/XMLSchema"

\quad xmlns:bans="http://example.wscs.com/2011/wsdl/BuyerAgent.wsdl"

\quad xmlns:bsns="http://example.wscs.com/2011/wsdl/BookStore.wsdl"

\quad xmlns:tns="http://example.wscs.com/2011/cdl/BuyerAgentAndBookStoreWSC"

\quad targetNamespace="http://example.wscs.com/2011/cdl/BuyerAgentAndBookStoreWSC"

\quad name="BuyerAgentAndBookStoreWSC"

\quad version="1.0"$\rangle$

\quad $\langle$informationType name="requestListofBooksType" type="bsns:tRequestListofBooks"/$\rangle$

\quad $\langle$informationType name="requestListofBooksResponseType"

\quad\quad type="bsns:tRequestListofBooksResponse"/$\rangle$

\quad $\langle$informationType name="listofBooksType" type="bsns:tListofBooks"/$\rangle$

\quad $\langle$informationType name="listofBooksResponseType"

\quad\quad type="bsns:tListofBooksResponse"/$\rangle$

\quad $\langle$informationType name="selectListofBooksType"

\quad\quad type="bsns:tSelectListofBooks"/$\rangle$

\quad $\langle$informationType name="selectListofBooksResponseType"

\quad\quad type="bsns:tSelectListofBooksResponse"/$\rangle$

\quad $\langle$informationType name="priceType" type="bsns:tPrice"/$\rangle$

\quad $\langle$informationType name="priceResponseType" type="bsns:tPriceResponse"/$\rangle$

\quad $\langle$informationType name="paysType" type="bsns:tPays"/$\rangle$

\quad $\langle$informationType name="paysResponseType" type="bsns:tPaysResponse"/$\rangle$

\quad $\langle$roleType name="BuyerAgent"$\rangle$

\quad\quad $\langle$behavior name="buyerAgent4BookStore" interface="bans:buyAgent4BookStoreInterface"/$\rangle$

\quad $\langle$/roleType$\rangle$

\quad $\langle$roleType name="BookStore"$\rangle$

\quad\quad $\langle$behavior name="bookStore4BuyerAgent" interface="rns:bookStore4BuyerAgentInterface"/$\rangle$

\quad $\langle$/roleType$\rangle$

\quad $\langle$relationshipType name="BuyerAgentAndBookStoreRelationship"$\rangle$

\quad\quad $\langle$roleType typeRef="tns:BuyerAgent" behavior="buyerAgent4BookStore"/$\rangle$

\quad\quad $\langle$roleType typeRef="tns:BookStore" behavior="bookStore4BuyerAgent"/$\rangle$

\quad $\langle$/relationshipType$\rangle$

\quad $\langle$choreography name="BuyerAgentAndBookStoreWSC"$\rangle$

\quad\quad $\langle$relationship type="tns:BuyerAgentAndBookStoreRelationship"/$\rangle$

\quad\quad $\langle$variableDefinitions$\rangle$

\quad\quad\quad $\langle$variable name="requestListofBooks" informationType="tns:requestListofBooksType"/$\rangle$

\quad\quad\quad $\langle$variable name="requestListofBooksResponse"

\quad\quad\quad\quad informationType="tns:requestListofBooksResponseType"/$\rangle$

\quad\quad\quad $\langle$variable name="listofBooks" informationType="tns:listofBooksType"/$\rangle$

\quad\quad\quad $\langle$variable name="listofBooksResponse" informationType="tns:listofBooksResponseType"/$\rangle$

\quad\quad\quad $\langle$variable name="selectListofBooks" informationType="tns:selectListofBooksType"/$\rangle$

\quad\quad\quad $\langle$variable name="selectListofBooksResponse"

\quad\quad\quad\quad informationType="tns:selectListofBooksResponseType"/$\rangle$

\quad\quad\quad $\langle$variable name="price" informationType="tns:priceType"/$\rangle$

\quad\quad\quad $\langle$variable name="priceResponse" informationType="tns:priceResponseType"/$\rangle$

\quad\quad\quad $\langle$variable name="pays" informationType="tns:paysType"/$\rangle$

\quad\quad\quad $\langle$variable name="paysResponse" informationType="tns:paysResponseType"/$\rangle$

\quad\quad $\langle$/variableDefinitions$\rangle$

\quad\quad $\langle$sequence$\rangle$

\quad\quad\quad $\langle$interaction name="InteractionBetweenBAandBS1"$\rangle$

\quad\quad\quad\quad $\langle$participate relationshipType="tns:BuyerAgentAndBookStoreRelationship"

\quad\quad\quad\quad\quad fromRoleTypeRef="tns:BuyerAgent" toRoleTypeRef="tns:BookStore"/$\rangle$

\quad\quad\quad\quad $\langle$exchange name="requestListofBooks"

\quad\quad\quad\quad\quad informationType="tns:requestListofBooksType" action="request"$\rangle$

\quad\quad\quad\quad\quad $\langle$send variable="cdl:getVariable('tns:requestListofBooks','','')"/$\rangle$

\quad\quad\quad\quad\quad $\langle$receive variable="cdl:getVariable('tns:requestListofBooks','','')"/$\rangle$

\quad\quad\quad\quad $\langle$/exchange$\rangle$

\quad\quad\quad\quad $\langle$exchange name="requestListofBooksResponse"

\quad\quad\quad\quad\quad informationType="requestListofBooksResponseType" action="respond"$\rangle$

\quad\quad\quad\quad\quad $\langle$send variable="cdl:getVariable('tns:requestListofBooksResponse','','')"/$\rangle$

\quad\quad\quad\quad\quad $\langle$receive variable="cdl:getVariable('tns:requestListofBooksResponse','','')"/$\rangle$

\quad\quad\quad\quad $\langle$/exchange$\rangle$

\quad\quad\quad $\langle$/interaction$\rangle$

\quad\quad\quad $\langle$interaction name="InteractionBetweenBAandBS2"$\rangle$

\quad\quad\quad\quad $\langle$participate relationshipType="tns:BuyerAgentAndBookStoreRelationship"

\quad\quad\quad\quad\quad fromRoleTypeRef="tns:BookStore" toRoleTypeRef="tns:BuyerAgent"/$\rangle$

\quad\quad\quad\quad $\langle$exchange name="sendListofBooks"

\quad\quad\quad\quad\quad informationType="tns:listofBooksType" action="request"$\rangle$

\quad\quad\quad\quad\quad $\langle$send variable="cdl:getVariable('tns:listofBooks','','')"/$\rangle$

\quad\quad\quad\quad\quad $\langle$receive variable="cdl:getVariable('tns:listofBooks','','')"/$\rangle$

\quad\quad\quad\quad $\langle$/exchange$\rangle$

\quad\quad\quad\quad $\langle$exchange name="sendListofBooksResponse"

\quad\quad\quad\quad\quad informationType="listofBooksResponseType" action="respond"$\rangle$

\quad\quad\quad\quad\quad $\langle$send variable="cdl:getVariable('tns:listofBooksResponse','','')"/$\rangle$

\quad\quad\quad\quad\quad $\langle$receive variable="cdl:getVariable('tns:listofBooksResponse','','')"/$\rangle$

\quad\quad\quad\quad $\langle$/exchange$\rangle$

\quad\quad\quad $\langle$/interaction$\rangle$

\quad\quad\quad $\langle$interaction name="InteractionBetweenBAandBS3"$\rangle$

\quad\quad\quad\quad $\langle$participate relationshipType="tns:BuyerAgentAndBookStoreRelationship"

\quad\quad\quad\quad\quad fromRoleTypeRef="tns:BuyerAgent" toRoleTypeRef="tns:BookStore"/$\rangle$

\quad\quad\quad\quad $\langle$exchange name="selectListofBooks"

\quad\quad\quad\quad\quad informationType="tns:selectListofBooksType" action="request"$\rangle$

\quad\quad\quad\quad\quad $\langle$send variable="cdl:getVariable('tns:selectListofBooks','','')"/$\rangle$

\quad\quad\quad\quad\quad $\langle$receive variable="cdl:getVariable('tns:selectListofBooks','','')"/$\rangle$

\quad\quad\quad\quad $\langle$/exchange$\rangle$

\quad\quad\quad\quad $\langle$exchange name="selectListofBooksResponse"

\quad\quad\quad\quad\quad informationType="selectListofBooksResponseType" action="respond"$\rangle$

\quad\quad\quad\quad\quad $\langle$send variable="cdl:getVariable('tns:selectListofBooksResponse','','')"/$\rangle$

\quad\quad\quad\quad\quad $\langle$receive variable="cdl:getVariable('tns:selectListofBooksResponse','','')"/$\rangle$

\quad\quad\quad\quad $\langle$/exchange$\rangle$

\quad\quad\quad $\langle$/interaction$\rangle$

\quad\quad\quad $\langle$interaction name="InteractionBetweenBAandBS4"$\rangle$

\quad\quad\quad\quad $\langle$participate relationshipType="tns:BuyerAgentAndBookStoreRelationship"

\quad\quad\quad\quad\quad fromRoleTypeRef="tns:BookStore" toRoleTypeRef="tns:BuyerAgent"/$\rangle$

\quad\quad\quad\quad $\langle$exchange name="sendPrice"

\quad\quad\quad\quad\quad informationType="tns:priceType" action="request"$\rangle$

\quad\quad\quad\quad\quad $\langle$send variable="cdl:getVariable('tns:price','','')"/$\rangle$

\quad\quad\quad\quad\quad $\langle$receive variable="cdl:getVariable('tns:price','','')"/$\rangle$

\quad\quad\quad\quad $\langle$/exchange$\rangle$

\quad\quad\quad\quad $\langle$exchange name="sendPriceResponse"

\quad\quad\quad\quad\quad informationType="priceResponseType" action="respond"$\rangle$

\quad\quad\quad\quad\quad $\langle$send variable="cdl:getVariable('tns:priceResponse','','')"/$\rangle$

\quad\quad\quad\quad\quad $\langle$receive variable="cdl:getVariable('tns:priceResponse','','')"/$\rangle$

\quad\quad\quad\quad $\langle$/exchange$\rangle$

\quad\quad\quad $\langle$/interaction$\rangle$

\quad\quad\quad $\langle$interaction name="InteractionBetweenBAandBS5"$\rangle$

\quad\quad\quad\quad $\langle$participate relationshipType="tns:BuyerAgentAndBookStoreRelationship"

\quad\quad\quad\quad\quad fromRoleTypeRef="tns:BuyerAgent" toRoleTypeRef="tns:BookStore"/$\rangle$

\quad\quad\quad\quad $\langle$exchange name="pays"

\quad\quad\quad\quad\quad informationType="tns:paysType" action="request"$\rangle$

\quad\quad\quad\quad\quad $\langle$send variable="cdl:getVariable('tns:pays','','')"/$\rangle$

\quad\quad\quad\quad\quad $\langle$receive variable="cdl:getVariable('tns:pays','','')"/$\rangle$

\quad\quad\quad\quad $\langle$/exchange$\rangle$

\quad\quad\quad\quad $\langle$exchange name="paysResponse"

\quad\quad\quad\quad\quad informationType="paysResponseType" action="respond"$\rangle$

\quad\quad\quad\quad\quad $\langle$send variable="cdl:getVariable('tns:paysResponse','','')"/$\rangle$

\quad\quad\quad\quad\quad $\langle$receive variable="cdl:getVariable('tns:paysResponse','','')"/$\rangle$

\quad\quad\quad\quad $\langle$/exchange$\rangle$

\quad\quad\quad $\langle$/interaction$\rangle$

\quad\quad $\langle$/sequence$\rangle$

\quad $\langle$/choreography$\rangle$

$\langle$/package$\rangle$

-------------------------------------------------------------------------------

At runtime, a WSC can be enabled in a centralized way or in a distributed way.The centralized way considers that the enablement of a WSC must be under supervision of an thirdly authorized party or all involved partners. This way maybe cause the supervisor becoming a performance bottleneck when bulk of interactions occur, but it can bring trustworthiness of interaction results if the supervisor is trustworthy itself. The distributed way argues that each WS interacts among others with constraints of a WSC and there is no need of a supervisor. But there do be cheating business behaviors of an intendedly \emph{incorrect}\footnote[1]{do not obey the constraints in a WSC} WS, that are unlike almost purely technical motivations of open Internet protocols.

At design time, the technical parts of a WSC can be generated automatically by use of the interface information of the partner WSes, such as WSDL descriptions and observable views of the WSOs. And more technical property descriptions can be included into a WSC, such as QoS agreements, security requirements, etc. Also as a business contract, the non-technical parts can be accomplished with human interferences, such as legal dispute processings.

\subsection{Relationship between Web Service Orchestration and Web Service Choreography}

\begin{figure}
  \centering
  %\vspace{5cm}
  \includegraphics{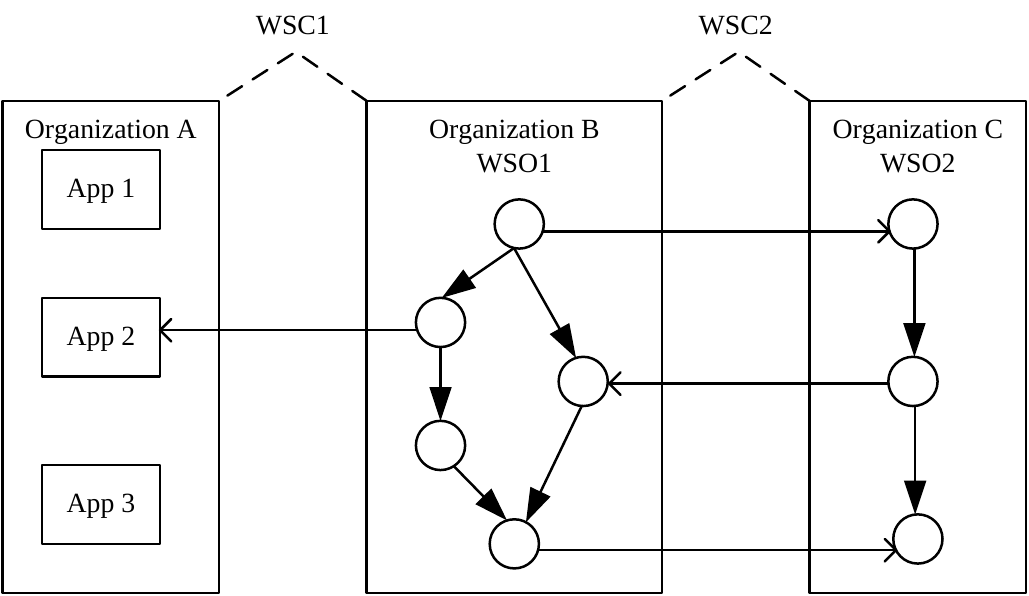}
  \caption{Relationship between WSO and WSC.}
  \label{Fig.WSOandWSC}
\end{figure}

A WSO orchestrates cross-organizational business activities implemented as WSes and can be accessed through its interfaces. And a WSC captures external interaction behaviors among its partner WSes through their interfaces. So the relationship between WSO and WSC are as Fig.\ref{Fig.WSOandWSC} illustrates.

Note that, a WSC need not be independent on WSOs.

\subsection{A Reference Model of Web Service Composition, RM-WSComposition}

\begin{figure}
  \centering
  %\vspace{5cm}
  \includegraphics{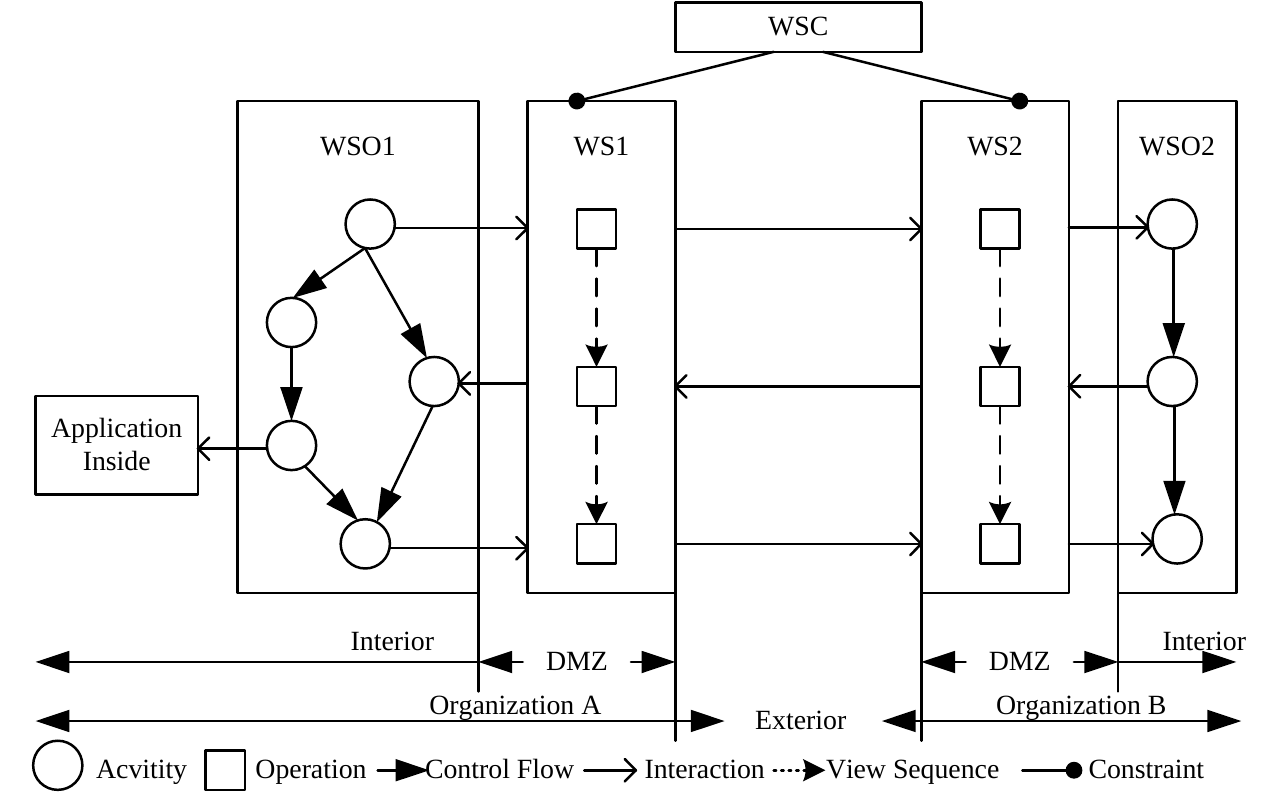}
  \caption{A Reference Model of Web Service Composition.}
  \label{Fig.RMComposition}
\end{figure}

Based on above discussions on WSO, interfaces of WSOs, WSC, and relationship between WSO and WSC, we get a reference model of WS Composition as Fig.\ref{Fig.RMComposition} illustrates.

At design time, according to some design methods, such as a bottom-up method\footnote[1]{firstly design inner WSOs, then get interfaces of WSOs, then generate a WSC} or a top-down method\footnote[2]{firstly design interfaces of WSOs, then refine the definitions of WSOs, then generate a WSC}, the elements of RMComposition including definitions of WSOs, interfaces of WSOs and a WSC are generated.

Then these definitions are deployed as Fig.\ref{Fig.RMComposition} shows. That is, WSOs are deployed in the interior of an organization, interfaces of WSOs are deployed in the DMZ of an organization, and a WSC may be enabled in a centralized way or a distributed way.

After the deployments, definitions of WSOs can be executed many times and each execution is called an WSO instance.  The interfaces of WSOs are always in ready for an incoming call. And mostly, a WSC is only running one time. That is, interactions among different WSO instances usually require under constraints of different WSCs, though a WSC may be reused for several times.

\subsection{BuyingBooks Example under RM-WSComposition}

The BuyingBooks example in section \ref{BuyingBooksExample} under RM-WSComposition illustrated in Fig.\ref{Fig.RMComposition} includes two WSOs (BuyerAgent and BookStore), four observable views (BuyerAgent4BookStore, BookStore4BuyerAgent, BookStore4RailwayCorp and BookStore4AirlineCorp),four WSes (BuyerAgent, BookStore, RailwayCorp and AirlineCorp) and three WSCs(BuyerAgentAndBookStore, BookStoreAndRailwayCorp and BookStoreAndAirlineCorp).

We define four WSes with WSDL, two WSes with WS-BPEL, four observable views with abstract process of WS-BPEL and three WSCs with WS-CDL.

\section{Issues on Formalization for Web Service Composition}

Based on RM-WSComposition, we discuss issues on formalization for WS Composition in this section.

\subsection{Formalization for Web Service Orchestration and Its Interface}

Formalization for WSO includes formalization for WSO, its interface, and the relationship between WSO and its interface.

\subsubsection{Formalization for Web Service Orchestration}

The first issue on formalization for WSO is formalizing control flows in a WSO. The control flows may be expressed in graphic model or structural model. Formalization for control flows may be based on different formal tools. By establishing the maps between control flows in a WSO and constructs in one formal model, control flows can be translated into expressions in the formal model. Then the properties, such as liveness and safety, can be verified.

Then, the following issues are formalization for technical mechanisms adopted in a WSO, such as compensation mechanism to support the so-called long running business transaction (LRBT), event processing mechanism, and exception throwing and processing mechanism, etc.

\subsubsection{Formalization for Interface of Web Service Orchestration}

The observable view of a WSO also includes control flows and inner transaction process mechanisms, so formalization for an observable view has the same issues as that of a WSO. Since the interface of a WSO includes the observable view of the WSO which is more expressive than the WSDL-expressed WS. That is, formalization for control flows and inner technical mechanisms may also be faced with.

\subsubsection{Formalization for Relationship between Web Service Orchestration and Its Interface}

A WSO interacts with outside via its interface, but, can the interface act in behalf of the inner WSO? That is, there must be a correct relationship between a WSO and its interface.

Since observable view of a WSO is an observable version of the inner WSO, formalization for relationship between a WSO and its interface can be based on some kind of equivalences, such as trace equivalence, bisimulation equivalence, and so on. Under such concepts of equivalence, a correct relationship between a WSO and its interface can be established. That is, an interface is \emph{same} as its inner WSO modulo such equivalence. Automatic generation of an interface from a inner WSO and also refinement of a inner WSO with an interface are also contents of a formal model.

\subsection{Formalization for Web Service Choreography}

Since WSC description language, such as WS-CDL, has not formal verification support. Formalization for WSC is to give WSC a formal semantics. Such an approach is also to establish maps between entities of a WSC description language and those of a formal tools to provide formal semantics support.

\subsection{Formalization for Relationship between Web Service Orchestration and Web Service Choreography}

Deos a WS can interact with another one? And also, does a WSO can interact with another one via their interfaces? Does the definition of a WSC compatible with its partner WSes or partner WSOs? To solve these problems, a correct relationship between WSO and WSC must be established.

The issues may include the compatibility verification of the interacting WSes or WSOs, the conformance verification of a WSC and its partner WSes or WSOs, automatic generation with correct assurance of a WSC from existing interface definitions of WSes or WSOs, refinement with correct assurance of interface definitions of WSes or WSOs from an existing WSC.

\section{Research Works on Formalization for Web Service Orchestration}

In this section, we will introduce the research works on formalization for WSO with a deep inspection about issues discussed in above section. Since Petri Net and Process Algebra are the two most influentially formal tools in formalizing WS Composition, we mainly introduce the works using these tools.

\subsection{Works Based on Petri Net}

Petri Net has good traditions in formalizing WSO and can be casted back to Aalst's WF-Net\cite{WF-Net}. \cite{FormBPELPNet} tries to give BPEL a formal semantics based on Petri Net. The works cover the standard behavior of BPEL, including atomic activities, control flows in term of structural activities, exception processing, event processing and LRBT mechanism.

\begin{figure}
  \centering
  %\vspace{5cm}
  \includegraphics{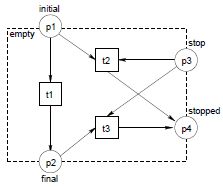}
  \caption{Transformation of Empty activity in BPEL into a Petri Net.}
  \label{Fig.EmptyActivity2PNet}
\end{figure}

Transforming every thing specified in BPEL into a construct of Petri Net is really a heavy work. The transformations of atomic activities include Empty activity, Receive activity, Wait activity, Assign activity, Reply activity, Invoke activity, Throw activity, and Terminate activity. For an example, the transformation of Empty activity is introduced as Fig.\ref{Fig.EmptyActivity2PNet} shows.

The transformations of control flows include Sequence activity, Flow activity, While activity, Switch activity, and Pick activity. And the transformations of inner transaction processing include Fault handler, Event handler and Compensation handler.

Thus a WSO described by BPEL can be transformed into a Petri Net entirely. Theories, such as WF-Net, and tools, such as WofBPEL\cite{WofBPEL} based on Petri Net can be applied for verification of liveness, safety and so on.

\cite{FormBPELPNet2} also transforms a WSO described by WS-BPEL into a so-called Service Workflow Net (SWN) which is a kind of colored Petri Net. And then it analyzes the compatibility of two services.

\subsection{Works Based on Process Algebra}

Process Algebra is also an influential tool in formalizing WS Composition. \cite{FormBPELPA} uses a Process Algebra CCS to model dynamic behaviors of a WS-BPEL business process.

Each message is represented by a CCS channel and atomic activities in BPEL are corresponding to input and output actions on these channels. The Flow activity is represented by parallel composition of CCS and the Sequence activity is also transformed using parallel composition. The Switch activity and Pick activity are modeled using choice operator in CCS. And Fault handler is also transformed into a process in CCS. Thus, a WSO described by WS-BPEL are transformed into a process in CCS.

For example, the BuyerAgent WSO in Fig.\ref{Fig.BuyingBooksExample} having the form:

$Seq=sequence\quad ReqLB\quad RecLB\quad SSB\quad RPB\quad PB$

is translated into a process in CCS:

$[\![Seq]\!](sq_0,sq_5)\|sq_0!().0$,

where
\begin{eqnarray}
[\![Seq]\!](sq_0,sq_5)&=&[\![ReqLB]\!](sq_0,sq_1)\nonumber\\
&\|&[\![RecLB]\!](sq_1,sq_2)\nonumber\\
&\|&[\![SSB]\!](sq_2,sq_3)\nonumber\\
&\|&[\![RPB]\!](sq_3,sq_4)\nonumber\\
&\|&[\![PB]\!](sq_4,sq_5)\nonumber
\end{eqnarray}

and $[\![P]\!](begin,end)=begin?().[\![P]\!](end)$.

Then by putting two processes of CCS which represent two WS-BPEL WSOs in parallel, compatibility of the two processes can be checked using theory of CCS.

A variant form of Process Algebra called Pi-Calculus is also used as the similar solution to formalizing WSO\cite{FormBPELPI}.

\subsection{Other Works}

Indeed, there are almost numerous research works on formalization for WSO and even workflow. To enumerate all the related works is unreachable. We give the representable works using different formal tools as follows. These works adopt the similar solution to giving WSO a formal foundation.

\cite{FormBPELWP} analyzes BPEL using a framework composed of workflow patterns and communication patterns and makes a comparison of BPEL, WSFL, XLANG, etc. \cite{FormBPEL2} formally defines an abstract executable semantics for BPEL based on an abstract state machine (ASM). \cite{FormBPEL3} transforms all primitive and structured activities of the BPEL into a composable Timed Automata (TA) for verification. \cite{FormWSOLOTOS} transforms all control flow patterns into LOTOS for validation. \cite{FormBPELModel} translates a WSO described by BPEL into an Extended Finite-state Automaton (EFA) and uses the SPIN model checker for the representation of the EFA model for verification. \cite{FormWSOCalculus} defines a calculus called Calculus for Orchestration of Web Services (COWS) to verify a WSO.

About the inner transaction processing mechanisms for WSO, there are some works on formalization for compensations. \cite{FormLongTransaction1} proposes a process compensation language called StAC and an operational semantics for StAC is introduced in \cite{FormLongTransaction2}. In \cite{FormLongTransaction3}, a trace semantics for LRBT is proposed.

About formalization for an interface of a WSO, especially an observable view, there are only a few works. \cite{InterorgaWorkfowPView1} and \cite{InterorgaWorkfowPView2} defines a process view for an cross-organizational workflow and discuss the usages of such a process view. But the model is almost a conceptual one and efforts still need to be done to comfort a WS Composition background. \cite{FormAbstractProcesses} points out that the abstract process in WS-BPEL does not prevent complex computations and is unnecessarily complex. They propose some restrictions on data manipulation constructs in an abstract process and also introduce a logic framework to verify such a restricted abstract process.

\subsection{Comparable Results on Related Works}

Through above introductions on formalization for WSO, we can see that there are many works on formalization WSO itself based on different formal tools and adopted similar ideas. Only a few works formalize the observable view of a WSO and almost no works give the observable view and a WSO a formal and natural relationship, such as automatic generation of a observable view from a inner WSO and refinement of a WSO from an existing observable view. In Table.\ref{Table.WSOWorks}, we conclude the related works on formalization of WSO.

\begin{table}
\centering
\begin{tabular}{@{}ll@{}}
Issues & Related Works\\
\hline
WSO Itself & \cite{FormBPELPNet} \cite{FormBPELPNet2} \cite{FormBPELPA} \\
&\cite{FormBPELPI} \cite{FormBPELWP} \\
&\cite{FormBPEL3} \cite{FormWSOLOTOS} \cite{FormBPELModel} \\
&\cite{FormWSOCalculus} \cite{FormBPEL2}\\

\hline
Inner Transaction Processing of WSO & \cite{FormLongTransaction1}\\
& \cite{FormLongTransaction2}\\
& \cite{FormLongTransaction3} \cite{FormLongTransaction4} \\
\hline
Interface and Observable View    &  \cite{InterorgaWorkfowPView1} \cite{InterorgaWorkfowPView2}\\ &\cite{FormAbstractProcesses}\\
\hline
Relationship between WSO and Its Interface    &  N/A \\
\end{tabular}
\caption{Works on Formalization for WSO.}
\label{Table.WSOWorks}
\end{table}

\section{Research Works on Formalization for Web Service Choreography}

In this section, we will introduce works on formalization for WSC. Also, we mainly introduce works based on Petri Net and Process Algebra.

\subsection{Works Based on Petri Net}

\cite{FormWSCPNet1} proposes a Petri Net approach for the design and analysis of WSC Using WS-CDL\cite{WS-CDL} as the language for describing WSC and Petri Net as a formalism to simulate and validate WSC. To capture timed and prioritized interactions, the Petri Net used is a kind of prioritized version of Time Petri Net (PTPN).

The methodology proposed in \cite{FormWSCPNet1} is called Collaborative Web Service Petri Nets (CWS-PNs) and consists of three phased: analysis, design and model validation. KAOS goal model is used in analysis phase to allow analysts to gather the requirements of WSCs in a hierarchical order, including time requirements, such as time-outs and other time constraints. After requirements are determined in analysis phase, then in design phase, a WS-CDL document is produced, that is, a WSC is generated. Finally, this WS-CDL document is transformed into a PTPN for verification and validation in model validation phase.

A PTPN semantics for WS-CDL is based on transformations of constructs of WS-CDL into those of a PTPN. Such transformations include basic activities, such as Interaction activity, Assign activity, Silent activity and NoAction activity, Workunits, ordering structures, such as Sequence, Parallel, and Choice, Exception blocks, Finalizer blocks, and Choreography composition.

\cite{FormWSCPNet2} also gives WSCI\cite{WSCI} a formal semantics based on Petri Net.

\subsection{Works Based on Process Algebra}

\cite{FormWSCPA} presents the semantics of WS-CDL in terms of Process Algebra CSP. Therefore, all the properties of a WSC to be checked can be verified with a CSP framework.

A CSP semantics for WS-CDL is also based on transformations of constructs of WS-CDL into those of CSP. The transformations include basic activities, such as Empty activity, Silent activity, Assign activity, Interaction activity and Throw activity, structural activities, such as Sequence activity, Parallel activity, Repetition Activity, and Choice activity, Exception handling and Compensation processing.

For example, the one-way Interaction activity is presented by the following CSP process:

$ch.op.r_f.r_t!In\rightarrow success\rightarrow Skip$

where a message $In$ from $r_f$ to $r_t$ is transmitted through channel $ch$ and $In$ is either the input message or the output message of operation $op$.

And the request-response Interaction activity is represented by the following CSP process:

$ch.op.r_f.r_t!In\rightarrow ch.op.r_t.r_f!Out\rightarrow success\rightarrow Skip$.

The BuyerAgentAndBookStore WSC in Fig.\ref{Fig.BuyingBooksExample} is represented by the following CSP process $P$:

\begin{eqnarray}
P &=& ch.requestListofBooks.BuyerAgent.BookStore!requestListofBooks\nonumber\\
&\rightarrow& ch.requestListofBooks.BookStore.BuyerAgent!requestListofBooksResponse \nonumber\\
&\rightarrow& ch.sendListofBooks.BookStore.BuyerAgent!listofBooks \nonumber\\
&\rightarrow& ch.sendListofBooks.BuyerAgent.BookStore!listofBooksResponse\nonumber\\
&\rightarrow& ch.selectListofBooks.BuyerAgent.BookStore!selectListofBooks\nonumber\\
&\rightarrow& ch.selectListofBooks.BookStore.BuyerAgent!selectListofBooksResponse \nonumber\\
&\rightarrow& ch.sendPrice.BookStore.BuyerAgent!price \nonumber\\
&\rightarrow& ch.sendPrice.BuyerAgent.BookStore!priceResponse\nonumber\\
&\rightarrow& ch.pays.BuyerAgent.BookStore!pays\nonumber\\
&\rightarrow& ch.pays.BookStore.BuyerAgent!paysResponse \nonumber\\
&\rightarrow& success\nonumber\\
&\rightarrow& Skip\nonumber
\end{eqnarray}

\cite{FormWSC1} formalizes WSCI based on Process Algebra CCS and can check whether two or more WSes are compatible.

\subsection{Other Works}

There are also many works to give WSC a formal semantics. \cite{FormWSCLogic} presents a formal model of WS-CDL based on a Spatio-Temporal Logic which can be used to reason on properties interested. \cite{FormWSCModel} proposes a denotational semantics model for WSCI. \cite{FormWSCValidation} develops a relational calculus to simulate and validate a WSC described by WS-CDL. \cite{WSCContract} relates theory of contracts and WSC with a notion of choreography conformance.

\section{Research Works on Formalization for Unifying Web Service Orchestration and Web Service Choreography}

WSO and WSC are two kinds composition patterns in WS Composition and have natural relationship under the requirements of cross-organizational business integrations. A formal framework covered WSO and WSC within which have a natural relationship between WSO and WSC is attractive and pursued. In this section, we will introduce works on formalization for unifying WSO and WSC.

\subsection{Works Based on Process Algebra}

\cite{WSOandWSCCSP} and \cite{WSOandWSCCSP2} use Process Algebra CSP as a formal basis for verifying the behavioral consistency among abstract and executable processes together with choreographic descriptions.

\begin{figure}
  \centering
  %\vspace{5cm}
  \includegraphics{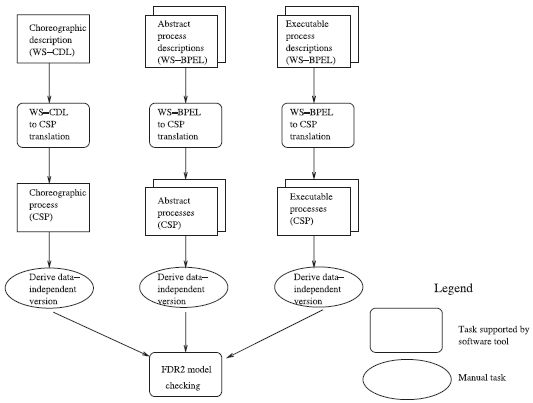}
  \caption{A CSP Approach to Unifying WSO and WSC.}
  \label{Fig.WSOandWSCCSP}
\end{figure}

As Fig.\ref{Fig.WSOandWSCCSP} shows, \cite{WSOandWSCCSP} and \cite{WSOandWSCCSP2} translate both constructs of WS-BPEL and constructs of WS-CDL into CSP processes. Then FDR2\cite{FDR2} is used to check behavioral consistency of WS-BPEL executable process, WS-BPEL abstract process and WS-CDL choreography.

\cite{WSOandWSC2}\cite{WSOandWSC3} transforms a WSO described by WS-BPEL and a WSC defined by WS-CDL both into two process algebra, and uses the form of bisimulation relation to define a notion of conformance between the WSO and the WSC.

\subsection{Works Based on Actors}

Actors\cite{Actor1}\cite{Actors} are basic concurrent computation models for distributed computing. \cite{WSOandWSCActors} uses actors as distributed objects\cite{ConcurrentObject} to unify WSO and WSC. A language called Actor-based Web Service Composition Language (Ab-WSCL) is designed for supporting RM-WSComposition illustrated in Fig.\ref{Fig.RMComposition}. More importantly, the semantics based on concurrent rewriting theory for actors is established. A notion called composibility is established between WSO and WS, WS and WS, WSO and WSO through interaction semantics analyses to give finely formal relations among these system components.

\cite{WSOandWSCActors} can be a system with formal semantics support, which not only can be used for verifications and simulations at design time, including verifying properties of WSO and WSC itself, and that of the relationship between WSO and WSC, but also can serve as a WS Composition runtime system with correctness assurance.

The UserAgent WSO in section \ref{BuyingBooksExample} programmed by Ab-WSCL is following.

---------------------------------------------------------------------------------------------------

WSO UserAgentWSO\{

\quad\quad AA requestLBAA

\quad\quad AA receiveLBAA

\quad\quad AA sendSBAA

\quad\quad AA receivePBAA

\quad\quad AA payBAA

\quad\quad WS ws-ref

\quad\quad List books

\quad\quad List selectedBooks

\quad\quad float prices

\quad\quad init(WS ws)\{

\quad\quad\quad\quad ws-ref := ws

\quad\quad\quad\quad requestLBAA := new RequestLBAA(self)

\quad\quad\quad\quad receiveLBAA := new ReceiveLBAA(self)

\quad\quad\quad\quad sendSBAA := new SendSBAA(self)

\quad\quad\quad\quad receivePBAA := new ReceivePBAA(self)

\quad\quad\quad\quad payBAA := new PayBAA(self)

\quad\quad\quad\quad other-local-computations

\quad\quad\}

\quad\quad requestLB() if true\{

\quad\quad\quad\quad other-local-computations

\quad\quad\quad\quad ws-ref $\leftarrow$ requestLB()

\quad\quad \}

\quad\quad receiveLB(List bs) if true\{

\quad\quad\quad\quad books := bs

\quad\quad\quad\quad other-local-computations

\quad\quad\quad\quad receiveLBAA $\leftarrow$ receiveLB(books)

\quad\quad \}

\quad\quad sendSB(sb) if true\{

\quad\quad\quad\quad selectedBooks := sb

\quad\quad\quad\quad other-local-computations

\quad\quad\quad\quad ws-ref $\leftarrow$ sendSB(selectedBooks)

\quad\quad \}

\quad\quad receivePB(float pb) if true\{

\quad\quad\quad\quad prices := pb

\quad\quad\quad\quad other-local-computations

\quad\quad\quad\quad receivePBAA $\leftarrow$ receivePB(prices)

\quad\quad \}

\quad\quad payB() if true\{

\quad\quad\quad\quad other-local-computations

\quad\quad\quad\quad ws-ref $\leftarrow$ payB()

\quad\quad \}

\}

---------------------------------------------------------------------------------------------------

Where requestLBAA, receiveLBAA, sendSBAA, receivePBAA, payBAA are activity actors (AAs). ws-ref is the reference of the interface WS of UserAgentWSO. Note that a WSO is always created by its interface WS in Ab-WSCL.

An AA is defined as a 8-tuples $(p,a,b,s,l,t,wso,ws)$. $p$ denotes a processing state of an actor. $a$ denotes an address of an actor, $b$ denotes a behavior of an actor, $s$ denotes a state of an actor, $l$ denotes the last event generated by an actor, $t$ denotes the transition map for an actor, $wso$ denotes AA's WSO, and $ws$ denotes WS of the WSO. An AA has the form of $p(a,b,wso,ws|\sigma:[s]\lambda:[l]\tau:[t])$ as a term.

$En_{ex}$ defines a predicate on actors, and $En_{ex}(a,s)$ holds if an actor with an a address $a$ and state $s$ is enabled for execution.

$recep$ defines receptionists of a fragment.

$acq$ defines acquaintances of a state or a value.

$a:a'\lhd v$ denotes a message with values $v$ sent from an actor with address $a'$ to an actor with address $a$.

$F$ denotes an actor fragment. $[F]\lceil \{a\}$ denotes a restriction of a fragment.

Semantics for UserAgentWSO and its AAs is following.

[\textbf{in}]
\\

$\langle F\rangle\rightarrow \langle F, a\lhd v\rangle$

\quad\quad\quad\quad if $F=[F', A]\lceil\{R\}$ and

\quad\quad\quad\quad $A=p(a,b,UserAgentWSO,ws|\sigma:[s]\lambda:[l]\tau:[t])$ and $a=UserAgentWSO$ and $a\in recep(F)$.
\\

[\textbf{out}]
\\

$\langle [F,a:a'\lhd v]\rangle\lceil\{R\}\rightarrow \langle [F]\lceil\{R\cup (acq(v)\cap recep(F))\}\rangle$

\quad\quad\quad\quad if $a\notin recep(F)$.
\\

[\textbf{create-RequstLBAA}]
\\

$A\rightarrow [A, A', WSO\blacktriangleleft(a',\textbf{ready},\{\})]\lceil\{a\}$

\quad\quad\quad\quad if $A=?(a,b_a,wso_a,ws_a|\sigma:[s_a]\lambda:[l_a]\tau:[t_a])$ and

\quad\quad\quad\quad $A'=?(a',b_{a'},wso_{a'},ws_{a'}|\sigma:[s_{a'}]\lambda:[\textbf{ready}]\tau:[WSO])$ and

\quad\quad\quad\quad $En_{ex}(a,s_a)$ holds and

\quad\quad\quad\quad $a'\notin acq(s)\cup\{a\}$

\quad\quad\quad\quad and $a=wso_a=wso_{a'}=WSO$ and $ws_a=ws_{a'}$ and $a'=RequstLBAA$.

Similarly to \textbf{create-RequstLBAA} rule, there are also \textbf{create-ReceiveLBAA} rule, \textbf{create-SendSBAA} rule, \textbf{create-ReceivePBAA} rule, and \textbf{create-PayBAA} rule.

\subsection{Other Works}

There are also some other works to unify WSO and WSC. \cite{WSOandWSCAutomata} uses Reo coordination language and constraint automata to derive a natural correspondence relationship between WSO and WSC. And also exception handling and finalization/compensation are used to connect a WSO and a WSC\cite{WSOandWSCException}.

\section{Conclusions and Trends}

In this section, we conclude the related research works and point out the trends on formalization for WS Composition.

\subsection{Conclusions}

As WS and WS Composition emerged at about ten years ago, there are plenty of works on formalization for WS Composition. As introductions above, the formalization works mainly focus on formalizing WSO itself and WSC itself to verify their properties based on various mathematical tools.

Other topics, such as interfaces of a WSO and their relationships, relationship between WSO and WSC, though are attractive, researchers still need to pay more attentions on them.

\subsection{Trends on Formalization for Web Service Composition}

\subsubsection{More Expressive Formal Models to Support Detailed Mechanisms in Web Service Composition}

One trend on formalization for WS Composition is that the research will be going deeper into the technical details adopted by WS Composition. For example, at beginning, formalization for WSO mainly focuses on control flow patterns and more technical mechanisms, such as compensation, transaction processing, exception handling, event handling are included in formal models later.

In future, we can see that more other technical mechanisms closed to the requirements of real cross-organizational applications will be formalized, such as security, QoS, etc.

\subsubsection{Unified Models to Support RM-WSComposition}

Another trend on formalization for WS Composition is that the research will cover more broader on aspects of WS Composition to fully support RM-WSComposition.

In future, we also can see that the unified models will emerge to support not only WSO itself, WSC itself, but also interfaces of a WSO and their relationships, relationship between WSO and WSC.

\subsubsection{Enforcement of the Formal Methods for Web Service Composition}

Practicability is every thing, so enforcement of the formal methods for WS Composition will be another trend. Now formal methods can be validated by using different tools, for examples, a Petri Net method can use WofBPEL\cite{WofBPEL}, a Process Algebra based method can use FDR2\cite{FDR2}.

Such methods and tools should be integrated into real WS Composition systems, including design time systems and runtime systems, for example, \cite{ToolofActiveBPEL} integrates a tool to support for BPEL verification in ActiveBPEL engine. Real WS Composition systems with inner formal models are also pursued in future, such as QoS-aware WSO engine with correctness assurance\cite{WSOEngineActors}.

%\newpage

\label{lastpage}

\end{document}